\documentstyle[floats,aps,psfig,preprint]{revtex}
\begin{document}
\newcommand{\slsh}{\not\!}
\newcommand{\kpp}{k_{++}}
\newcommand{\skpp}{\!\not\!k_{++}}
\newcommand{\kpm}{k_{+-}}
\newcommand{\skpm}{\!\not\!k_{+-}}
\newcommand{\kmp}{k_{-+}}
\newcommand{\skmp}{\!\not\!k_{-+}}
\newcommand{\kop}{k_{0+}}
\newcommand{\kopm}{k_{0+\mu}}
\newcommand{\ga}{\gamma_\alpha}
\newcommand{\gb}{\gamma_\beta}
\newcommand{\gm}{\gamma_\mu}
\newcommand{\svkpp}{\sigma_V(k_{++}^2)}
\newcommand{\sskpp}{\sigma_S(k_{++}^2)}
\newcommand{\svkmp}{\sigma_V(k_{-+}^2)}
\newcommand{\sskmp}{\sigma_S(k_{-+}^2)}
\newcommand{\svkpm}{\sigma_V(k_{+-}^2)}
\newcommand{\sskpm}{\sigma_S(k_{+-}^2)}
\title{Dipole Moments of the Rho Meson}
\author{M.B. Hecht, B.H.J. McKellar}
\date{\today}
\draft
\preprint{UM-P-97/15}
\address{{\it School of Physics, University of Melbourne, Parkville VIC 3052,
     Australia}}
\maketitle
\begin{abstract}
The electric and magnetic dipole moments of the rho meson are 
calculated using the propagators and vertices derived from the
QCD Dyson-Schwinger equations. Results obtained from using the 
Bethe-Salpeter amplitude studied by Chappell, 
Mitchell and Tandy, and
Pichowsky and Lee, are compared.
The rho meson EDM is generated
through the inclusion of a quark electric dipole moment, which
is left as a free variable.
These results are compared to the 
perturbative results to obtain a measure of the effects of quark
interactions and confinement. The two dipole moments are also calculated using
the phenomenological MIT bag model to provide a further basis for 
comparison.
\end{abstract}
\newpage

\section{Introduction}
The electric and magnetic dipole moments of the rho meson, the lowest lying
vector particle state are calculated using the  semi-phenomenological n-point
functions of Quantum Chromodynamics (QCD) obtained in the 
Dyson-Schwinger/Bethe-Salpeter framework. The rho meson is studied to
provide some insight into the effect of QCD and confinement
on the electromagnetic moments of hadrons, especially the question of the
existence of a non-zero EDM for the neutron and the applications this has
to the study of CP violation. In practice one is interested in these
parameters for the nucleon and other hadrons, where they may be measured.
But the study of the rho allows us to obtain insights without venturing into
the complications of the three body system in QCD.

The Dyson-Schwinger equations are a series of coupled integral equations
relating the n-point functions of QCD to one another. They provide a 
continuum method of calculation in the strong coupling regime of QCD and as
such are able to relate the Greens functions of QCD to hadronic properties.
For a review of this topic see \cite{RobertsWilliams94}.
The use of the Dyson-Schwinger and Bethe-Salpeter equations to calculate
hadronic properties has been applied to pion
observables \cite{AlkoferBenderRoberts95,Roberts95,Roberts96,FrankRoberts96},  
kaon electromagnetic form factors
\cite{BurdenRobertsThomson96}, $\rho$-$\omega$ mixing 
\cite{MitchellTandy97} and the anomalous $\gamma\pi^*
\rightarrow\pi\pi$ form factor \cite{AlkoferRoberts96}.

This calculation makes use of a model form for the quark propagator which
ensures confinement by having no Lehmann representation, preventing free
quark production thresholds, {\it i.e.,} the quark cannot go ``on mass-shell''.
A phenomenological form for the rho amplitude, which models the
momentum distribution of the
quarks inside, and the finite size of, the rho, is used, as well as the
Ball-Chiu/Curtis-Pennington quark-photon vertex which ensures that both
the Ward-Takahashi Identity and multiplicative renormalizability are
preserved. 
The CP violating electric dipole term is introduced via
a quark EDM which is left as a free parameter which can be varied.

Results obtained from using the Bethe-Salpeter amplitude studied by
Mitchell and Tandy \cite{MitchellTandy97}, Chappell \cite{Chappell96},
and Pichowsky and Lee \cite{PichowskyLee96}, are compared.
Perturbation theory results \cite{Pallaghy96}, which ignore the 
dressing of the vertex functions
and propagators, are also included 
to observe the effect of
using a quark propagator that correctly incorporates confinement and of 
modelling the dressed quark core of the rho.

A phenomenological model, the MIT bag model is also examined. This model
incorporates relativistic, point-like confined quarks but  ignores one-gluon
exchange and centre of mass motion. Both the magnetic and electric dipole 
moments are calculated in the bag model providing a basis for comparison
between the Dyson-Schwinger/Bethe-Salpeter framework and a purely
phenomenological model, one that has been well studied over the years.

\section{The Dipole Moments}
\label{sec2}
\subsection{The $\rho\rho\gamma$ Vertex}
The form for the CP conserving $\rho\rho\gamma$ vertex is 
\cite{SalamDelbourgo64},
\begin{eqnarray}
d_\alpha^{\ \nu} (p)\Gamma_{\mu\nu\sigma}d_\beta^{\ \sigma} (p^\prime ) & = & 
d_\alpha^{\ \nu} (p)[ (p+p^\prime )_\mu (-g_{\nu\sigma}{\cal E}(q^2)
    \nonumber \\
{} & {} & +q_\nu q_\sigma {\cal Q}(q^2)) + \left(g_{\mu\nu}q_\sigma - 
        g_{\mu\sigma}q_\nu \right){\cal M}(q^2)]
        d_\beta^{\ \sigma} (p^\prime ), 
\end{eqnarray}
where $d_{\alpha\beta}$ are the rho polarisation vectors,
\begin{equation}
d_{\alpha\beta}(k) = g_{\alpha\beta} - \frac{k_\alpha k_\beta }{k^2} .
\end{equation}
The form factors ${\cal E,Q}$ and ${\cal M}$ reduce in the limit as
$q^2\rightarrow 0$ to
\begin{eqnarray}
{\cal E}(0) & = & 1, \nonumber \\
{\cal M}(0) & = & \mu,\mbox{ the magnetic moment in units of }e/2m \nonumber \\
{\cal Q}(0) & = & \frac{2}{m^2}\left(Q+\mu -1\right),\mbox{  where Q is the
                   quadrupole moment in units of } e/m^2 \nonumber
\end{eqnarray}

\subsection{Projecting Out Charge and Dipole Terms}
The integrals for the rho dipole moments that will be derived later, 
equations (10) and
(39), are too difficult to perform analytically. This means that to isolate
relevant terms they will have to be projected out from the integral before it is calculated numerically.   Defining
\begin{equation}
V_{\alpha\mu\beta} = d_\alpha^{\ \nu}(p)\Gamma_{\nu\mu\sigma}d_\beta^{\ \sigma}
                       (p^\prime),
\end{equation}
and with the projection operator
\begin{equation}
P_{1}^{\alpha\mu\beta} = d^\alpha_{\ \nu^\prime}(p)P^\mu
                           (-g^{\nu^\prime\sigma^\prime})d^\beta_{\ 
                           \sigma^\prime}(p^\prime),
\end{equation}
where $P_\mu = (p+p^\prime)_\mu$, the following result is obtained,
\begin{equation}
P_{1}^{\alpha\mu\beta}V_{\alpha\mu\beta} \stackrel{\lim q^2
           \rightarrow 0}{\rightarrow}12{\cal E}(0).
\end{equation}
To project out the magnetic dipole term we use the following projection
operator,
\begin{equation}
P_{2}^{\alpha\mu\beta} = d^\alpha_{\ \nu^\prime}(p)
     \left[\frac{g^{\mu\nu^\prime}
     q^{\sigma^\prime}-g^{\mu\sigma^\prime}q^{\nu^\prime}}{q^2}+
     \frac{P^\mu g^{\nu^\prime\sigma^\prime}}{6p^2}\right]
     d^\beta_{\ \sigma^\prime}(p^\prime),
\end{equation}
which is such that
\begin{equation}
P_2^{\alpha\mu\beta}V_{\alpha\mu\beta}\stackrel{\lim q^2\rightarrow 0}{
     \rightarrow}4{\cal M}(0).
\end{equation}

\subsection{Quark Contribution to the Magnetic Dipole Moment}
All calculations below, except those involving the bag model,
are carried out in Euclidean space where
$g_{\mu\nu} = \delta_{\mu\nu}$, $\gamma_\mu = \gamma_\mu^\dagger$ and
$\left\{ \gamma_\mu,\gamma_\nu\right\} = 2\delta_{\mu\nu}$.

The impulse approximation to the quark contribution to the
magnetic dipole moment is given by (see Fig.~\ref{figure2})
\begin{equation}
M_{fi} = \sum^{}_{\rm flavour=u,d}\varepsilon^\alpha (p)\varepsilon^\mu (q)
     \varepsilon^\beta (p^\prime)eQ_{\rm flavour}I^{\rm flavour}_{
     \alpha\mu\beta}.    
\end{equation}
$Q_{\rm flavour}$ is the charge of the quark interacting with the photon.
The $u$ and $d$-quarks are treated as identical except for their charge and so
the contribution becomes
\begin{equation}
M_{fi} = \varepsilon^\alpha (p)\varepsilon^\mu (q)\varepsilon^\beta (p^\prime)e
     I_{\alpha\mu\beta},
\end{equation}
\begin{eqnarray}
I_{\alpha\mu\beta} & = & (-1)\int^{}_{}\frac{d^4k}{(2\pi)^4}tr_{CFD}[
     \Gamma^\rho_\beta (k;p-q)S(k_{-+})\imath\Gamma^{BC+CP}_
     \mu (k_{++},k_{-+}) 
     \nonumber \\
                   &   &\times S(k_{++})\Gamma^\rho_\alpha (k+q/2;p)
     S(k_{+-})].
\end{eqnarray}
Where $k_{\alpha\beta}=k+\frac{\alpha q}{2}+\frac{\beta p}{2}$, $\Gamma^\rho$
refers to the rho meson amplitude and $\Gamma^{BC+CP}_{\mu}$ is the dressed
quark-photon vertex, and $S(p)$ is the dressed quark propagator for a quark
of momentum $p$, which will all be discussed in detail below.

\subsection{Quark Propagators}
The general form for the solution to the quark propagator
Dyson-Schwinger equation \cite{RobertsWilliams94} is
\begin{eqnarray}
S(p) & = & -\imath \slsh p\sigma_V(p^2)+\sigma_S(p^2) \nonumber \\
     & = & \left(\imath \slsh p A(p^2)+B(p^2)\right)^{-1}.
\end{eqnarray}
A model form for the propagator is given by \cite{Roberts95,Roberts96}
\begin{eqnarray}
\bar{\sigma}_S (x) & = & C_{\bar{m}}e^{-2x}+\left(\frac{1-e^{-b_1x}}{b_1x}
     \right)\left(\frac{1-e^{-b_3x}}{b_3x}\right)\left(b_0+b_2\left(
     \frac{1-e^{-\Lambda x}}{\Lambda x}\right)\right) \nonumber \\
                   &   & +\frac{\bar{m}}{x+\bar{m}^2}
     \left(1-e^{-2(x+\bar{m}^2)}\right), \\
\bar{\sigma}_V(x)  & = & \frac{2(x+\bar{m}^2)-1+e^{-2(x+\bar{m}^2)}}
     {2(x+\bar{m}^2)^2}-\bar{m}C_{\bar{m}}e^{-2x},
\end{eqnarray}
where $x=p^2/2D,$ $\bar{\sigma}_V(x) = (2D)\sigma_V(p^2),$
$\bar{\sigma}_S = \sqrt{2D}\sigma_S(p^2)$ and $\bar{m} = m/\sqrt{2D},$ $D$
is a mass scale. ($\Lambda = 10^{-4}$ is chosen to decouple the small
and large ${\rm spacelike-}p^2$ behaviour in equation (12); {\it i.e.,}
to allow for $b_0$ to govern the ultraviolet behaviour and $b_2$
the infrared.)
The parameters $C_{\bar{m}}$, $\bar{m}$, $b_0,\ldots,b_3$ are 
\cite{BurdenRobertsThomson96}
\begin{equation}
\begin{array}{|ccc|} \hline
C_{\bar{m}\ne 0} & = & 0.0  \\ \hline
C_{\bar{m}=0} & = & 0.121  \\ \hline
\bar{m}     & = & 0.00897  \\ \hline
b_0         & = & 0.131  \\ \hline
b_1         & = & 2.90  \\ \hline
b_2         & = & 0.603  \\ \hline
b_3         & = & 0.185  \\ \hline
\end{array}
\end{equation}
with the mass scale $D=0.160\;{\rm GeV}^2$ chosen to give the correct
value for $f_\pi$. 
This form for the quark propagator is based upon studies of the Dyson-Schwinger 
equation for $S(p)$ using a gluon propagator with an infrared singularity,
\begin{equation}
g^2D_{\mu\nu}(k) \equiv \left(\delta_{\mu\nu}-\frac{k_\mu k_\nu}{k^2}\right)
  8\pi^4D\delta^4(k),
\end{equation}
and a dressed quark-gluon vertex
\cite{Roberts96,BurdenRobertsThomson96}. Quark confinement is characterised
by the non-observation of
free quark states. A pole in the quark propagator in the
timelike region would be a sign of a free quark state, just as a 
free bound state
is manifest as a pole in a n-point Green's function. 
Thus a sufficient condition
for the lack of free quark production thresholds is the absence of timelike
poles in the propagator. The model quark propagator given 
above is an entire function
(except at timelike-$p^2 = \infty$) and so does not have a Lehmann 
representation. This means it can be interpreted as describing a 
confined particle and it
ensures the lack of the unphysical singularities corresponding to
free quarks in $I_{\alpha\mu\beta}$.

The dipole moments are also calculated using a form for the quark
propagator developed by Mitchell and Tandy
\cite{MitchellTandy97}, to investigate $\rho$-$\omega$ mixing.
This propagator is given by
\begin{equation}
\bar{\sigma}_S(x) = C_{\bar{m}}e^{-2x}+\frac{\bar{m}}{x}\left( 1-e^{-2x}
\right) ,
\end{equation}
\begin{equation}
\bar{\sigma}_V(x) = \frac{e^{-2x}-(1-2x)}{2x^2}-\bar{m}C_{\bar{m}}e^{-2x}.
\end{equation}
To fix the parameters $\lambda = \sqrt{2D}$ and $C_{\bar{m}}$, 
a fit to $\langle \bar{q}q
\rangle$, $f_\pi$, $r_\pi$ and the $\pi$-$\pi$ scattering lengths was done.
With $\frac{1}{2}(m_u+m_d) = 16$ MeV a best fit was obtained for
$\lambda = 0.889$ GeV and $C_{\bar{m}} = 0.581$
\cite{MitchellTandy97}.
This form has a deficiency which can be seen in it's failure to correctly
model the behaviour of $\sigma_S$ away from $x = 0$, in the massless limit
with dynamically broken chiral symmetry \cite{Roberts96}.
The large value for the mass of the quark used is related to the deficiency
of the propagator mentioned above \cite{Mitchell97}.
\subsection{Rho-meson Amplitude}
The dominant Bethe-Salpeter amplitude for the rho meson is given by
\cite{PraschifkaCahillRoberts89,JainMunczek93},
\begin{equation}
\Gamma^{l}_{\rho\;\mu} (k,p) = \imath \left(\gamma_\mu -\frac{\slsh p p_\mu}
     {p^2}\right)\tau^l\frac{\Gamma_\rho (k,p)}{N_\rho},
\end{equation}
where $k$ is the relative momentum of the quark and anti-quark, $p$
is the momentum of the rho meson and $l$ and $\mu$ are flavour and
Dirac indices respectively.
This form ignores other allowable Dirac structure in the
vector meson Bethe-Salpeter amplitude and so introduces
errors of the order of 10\% \cite{FrankRoberts96}. Using the quark propagator
defined in equations (12) and (13) and a Ball-Chiu quark-photon vertex (see 
later), Chappell uses the following approximate form for $\Gamma_\rho (k,p)$ 
\cite{Chappell96}
\begin{equation}
\Gamma_\rho = e^{-k^2/a_1^2}+\frac{a_2}{1+\frac{k^2}{\alpha a_1}},
\end{equation}
with $a_1 = 0.38845$, $a_2 = 0.01478, \alpha = 2$. The values for the 
parameters were found by fitting to the experimental values
of $f_\rho$ and $g_{\rho\pi\pi}
$ \cite{Chappell96}.
Pichowsky and Lee, also using the quark propagator defined in equations
(12) and (13) and a quark-photon vertex of the Ball-Chiu type, use an
identical form for $\Gamma_\rho$, given by \cite{PichowskyLee96}
\begin{equation}
\Gamma_\rho (k,p) = e^{-k^2/a_V^2}+\frac{c_V}{1+k^2/b_V^2}.
\end{equation}
Fitting to the experimental values for $f_\rho$ and $g_{\rho\pi\pi}$
they obtain
values for the parameters of $a_V = 0.400$, $b_V = 0.008$ and $c_V = 125.0$.
Using the quark propagator defined in equations (15) and (16) Mitchell
and Tandy use a form for the amplitude $\Gamma_\rho$ given by
\begin{equation}
\Gamma_\rho (k,p) = e^{-k^2/a^2},
\end{equation}
where $a = 0.194$ GeV.
The normalisation for
the rho amplitude is fixed by \cite{PichowskyLee96}
\begin{eqnarray}
p_\mu \left(\delta_{\alpha\beta}-\frac{p_\alpha p_\beta}{p^2}\right) & = &
	N_c{\rm tr}_D\int_{}^{}\frac{d^4k}{(2\pi )^4}
	\frac{\partial S(k_+)}{\partial p_\mu}\Gamma_{\rho\alpha}(k,p)S(k_-)
	\Gamma_{\rho\beta}(k,p) \\ \nonumber
 &  & + N_c{\rm tr}_D\int_{}^{}\frac{d^4k}{(2\pi )^4}
	S(k_+)\Gamma_{\rho\alpha}(k,p)
        \frac{\partial S(k_-)}{\partial p_\mu}\Gamma_{\rho\beta}(k,p),
\end{eqnarray}
where $k_{\alpha} = k+\frac{\alpha p}{2}$. This condition,
along with the fact that the quark-photon vertex given below
obeys the Ward Identity, ensures that
${\cal E}(q^2 = 0) = 1$, {\it i.e.}, that the rho has unit charge
\cite{Roberts96}.

Using the following transformation properties: $S^T(-k) = C^\dagger S(k)C$;
$\Gamma_\alpha^{\rho T}(-k,p) = -C^\dagger \Gamma_\alpha^\rho (k,p)C$
and $\Gamma_\mu^{{\rm BC+CP}\ T}(-p,-q) = 
-C^\dagger\Gamma_\mu^{{\rm BC+CP}}(p,q)C$, where $C = \gamma_2\gamma_4$ is
the charge conjugation operator, and the Ward-Takahashi Identity,
one can show that current conservation
holds \cite{Roberts96}, {\it i.e.},
\begin{equation}
q_\mu I_{\alpha\mu\beta} = 0.
\end{equation}

Given a description for the quark-photon vertex, the dipole moments of the
rho can now be calculated using equations (12), (13) and (18); (12), (13) and
(19), and (15), (16) and (20).

\subsection{Quark-Photon Vertex}
The quark-photon vertex also satisfies its own Dyson-Schwinger equation,
but solving this integral equation is difficult.
Despite this, a realistic ansatz for the vertex function has been developed
by constraining its form using certain criteria
\cite{BurdenRobertsWilliams93,DongMunczekRoberts94}.
The quark-photon vertex ansatz thus obtained is given by 
\begin{equation}
\Gamma_\mu^{{\rm BC+CP}}(p,q) = \Gamma_\mu^{{\rm BC}}(p,q) + 
     \Gamma_\mu^{{\rm CP}}(p,q).
\end{equation}
The Ball-Chiu vertex, $\Gamma_\mu^{{\rm BC}}$ has the form \cite{BallChiu80}
\begin{eqnarray}
\Gamma_\mu^{{\rm BC}}(p,q) & = & \frac{A(p^2)+A(q^2)}{2}\gamma_\mu +
     \frac{(p+q)_\mu}{p^2-q^2}\left[\frac{1}{2}(A(p^2)-A(q^2))(\slsh p +
     \slsh q)\right. \nonumber \\
                      &   & -\left.\frac{}{}\imath (B(p^2)-B(q^2))\right]. 
\end{eqnarray}
This vertex ansatz is completely described by the dressed quark propagator
and satisfies both the Ward-Takahashi and Ward Identities, is free
of kinematic singularities as $q^2\rightarrow p^2$, transforms correctly
under appropriate transformations and reduces to the perturbative limit.
To ensure multiplicative renormalizability Curtis and
Pennington added a transverse piece to the Ball-Chiu vertex
\cite{CurtisPennington90}. This term has the form given below
\begin{equation}
\Gamma_\mu^{{\rm CP}}(p,q) = \left(\frac{-\imath\gamma_\mu (p^2-q^2)-(p+q)_\mu
     (\slsh p-\slsh q)}{2d(p,q)}\right)\left(A(p^2)-A(q^2)\right),
\end{equation}
with
\begin{equation}
d(p,q) = \frac{1}{p^2+q^2}\left[ (p^2-q^2)^2+\left( M^2(p^2)+M^2(q^2)\right)^2
     \right],
\end{equation}
\begin{eqnarray}
M(p^2) = \frac{B(p^2)}{A(p^2)}. \nonumber
\end{eqnarray}

\subsection{The Magnetic Dipole Moment}
The dipole moment can now be calculated. The colour, flavour and Dirac 
traces are performed for the integral $I_{\alpha\mu\beta}$, given in
appendix A,  and
the charge and dipole terms projected out using the projection operators
outlined above. The integrals are performed and the dipole calculated using
\begin{eqnarray}
\mu & = & \, \lim_{q^2\rightarrow 0}{\cal M}(q^2)\nonumber \\
    & = & \, \lim_{q^2\rightarrow 0}P_2^{\alpha\mu\beta}
          I_{\alpha\mu\beta}/4.
\end{eqnarray}

The calculation is performed in the Breit Frame where
\begin{eqnarray}
q & = & (0,0,Q,0), \nonumber \\
p & = & \left( 0,0,Q/2,\imath\sqrt{m_\rho^2+Q^2/4}\right), \nonumber \\
p^\prime & = & \left( 0,0,-Q/2,\imath\sqrt{m_\rho^2+Q^2/4}\right). \nonumber
\end{eqnarray}
A transformation to 4-dimensional hyper-spherical coordinates is carried out
so that
\begin{equation}
k = \left|k\right| (\sin\beta\sin\theta\cos\phi ,\sin\beta\sin\theta\sin\phi 
    ,\sin\beta\cos\theta ,\cos\beta ), \nonumber
\end{equation}
with
\begin{equation}
\int_{}^{}d^4k = \int_0^\infty k^3dk\int_0^\pi \sin^2\beta d\beta
                 \int_0^\pi\sin\theta d\theta\int_0^{2\pi}d\phi. \nonumber
\end{equation}
There is no $\phi$ dependence in the integrand so the $\phi$ integral
is performed analytically, contributing a factor of $2\pi$.

The integral is then performed numerically using Gaussian quadrature methods
to obtain the results given in Table~\ref{table1}. The results obtained
using the framework of Chappell, and Pichowsky and Lee, are within $10\%$
of each other which is indicative of the similarities of the approaches.
The perturbative result
from \cite{Pallaghy96}, and the non-relativistic SU(6) result are given
for comparison. The later result is simply \cite{Hecht94},
\begin{equation}
\mu_{\rho^+} = \frac{e}{2m_q}.
\end{equation}
In equation (31) the quark mass is to be interpreted a constituent
quark mass of say 340 MeV \cite{Close79}, which gives $\mu_{\rho^+} = 
2.26 \ e/2m_\rho$. Thus one could say that the effect of 
QCD as interpreted in our 
calculation is to convert $m_q$ from the current quark value to an
appropriate constituent quark value. Note that the improper use of a current
quark mass in equation (31) gives the large value $\mu_{\rho^+} = 152\ 
e/2m_\rho$.

\subsection{The Electric Dipole Moment}
CP violation is a feature of the weak interaction which is not well
understood at present. It will, in principle, give rise to electric dipole
moments for quarks and hadrons. In calculating hadronic electric dipole
moments, the question of the effects of confinement and QCD in going from
the quark EDM to the hadron EDM have not been studied. Here we study these
effects for the rho meson.

The magnetic dipole of the rho meson comes from the term in the 
Salam-Delbourgo vertex function
\begin{eqnarray*}
(g_{\alpha\mu}q_\beta -g_{\beta\mu}q_\alpha ){\cal M}.
\end{eqnarray*}
This results in a term in the $\rho\rho\gamma$ matrix element that looks like
\begin{eqnarray*}
\varepsilon (p).F.\varepsilon (p^\prime ){\cal M},
\end{eqnarray*}
where
\begin{eqnarray*}
F_{\alpha\beta} = \partial_\beta A_\alpha -\partial_\alpha A_\beta
\end{eqnarray*}
is the electromagnetic field tensor. To find the vertex function term that
contributes to the electric dipole moment the familiar correspondence is used:
\begin{eqnarray}
F_{\alpha\beta} & \rightarrow & \tilde{F}_{\alpha\beta} = \frac{1}{2}
      \varepsilon_{\alpha\beta\mu\nu}F^{\mu\nu} \nonumber \\
{\cal M}(q^2)   & \rightarrow & {\cal D}(q^2),
\end{eqnarray}
where ${\cal D}(0) = d$, the electric dipole moment in units of e/2m.
Thus as 
\begin{eqnarray}
\frac{1}{2}\varepsilon_{\alpha\beta\mu\nu}F^{\mu\nu} = 
      \varepsilon_{\alpha\beta\mu\nu}A^\mu q^\nu, \nonumber 
\end{eqnarray}
the term in the $\rho\rho\gamma$ vertex that defines the rho EDM is
\begin{equation}
\varepsilon_{\alpha\beta\mu\nu}q^\nu{\cal D}(q^2).
\end{equation}

Once again the electric and dipole terms will need to be projected out. The
vertex function is given by
\begin{eqnarray}
V^{{\rm edm}}_{\alpha\mu\beta} & = & d_\alpha^{\ \nu}(p)\Gamma_{\nu\mu\sigma}
      d_\sigma^{\ \beta}(p^\prime) \nonumber \\
                              & = & d_\alpha^{\ \nu}(p)\left[ P_\mu 
      (-g_{\nu\sigma}{\cal E}(q^2)+q_\nu q_\sigma{\cal Q}(q^2))\right.
      \nonumber \\
                              &   & +(g_{\mu\nu}q_\sigma -g_{\mu\sigma}q_\nu )
      {\cal M}(q^2)
      +\varepsilon_{\nu\sigma\mu\rho}q^\rho
      \left.{\cal D}(q^2)\right] d_\beta^{\ \sigma}(p^\prime ).
\end{eqnarray}
Using a projection operator of the form
\begin{equation}
P_3^{\alpha\mu\beta} = -\left( \frac{1}{2q^2}\right) d^{\alpha\nu^\prime}(p)
    \varepsilon_
    {\nu^\prime\sigma^\prime}^{\ \ \ \ \mu\rho}q_\rho d^{\beta\sigma^\prime}(
    p^\prime),
\end{equation}
the result 
\begin{equation}
P_3^{\alpha\mu\beta}V_{\alpha\mu\beta}\stackrel{\lim q\rightarrow 0}
{\rightarrow}
{\cal D}(0)
\end{equation}
follows.

\subsection{Quark Contribution to the Electric Dipole Moment}
To introduce an electric dipole moment in the impulse approximation
the $qq\gamma$ vertex needs to be modified to introduce a CP-violating
term. Thus the following vertex is used:
\begin{equation}
\Gamma_\mu (p,q) = \Gamma_\mu^{{\rm BC+CP}}(p,q)-
	\imath D_q\gamma_5
	\sigma_{\mu\rho}(p-q)^\rho,
\end{equation}
where $D_q$ is the EDM of the quark. This form for the $qq\gamma$ vertex
still ensures that both the Ward-Takahashi and Ward Identities are preserved,
is free of kinematic singularities as $q^2\rightarrow p^2$ and reduces to
the perturbative limit. It also still preserves 
multiplicative renormalizability. However we have ignored any structure of the 
CP-violating term which may be generated by strong or electromagnetic
interactions.

Following the same procedure as outlined above, the impulse
approximation to the EDM is given by
\begin{equation}
{\cal M}_{fi} = \varepsilon^\alpha (p)\varepsilon^\mu (q)\varepsilon^\beta
                (p^\prime )eI^{\rm edm}_{\alpha\mu\beta},
\end{equation}
where the integral $I^{\rm edm}_{\alpha\mu\beta}$ is given by
\begin{eqnarray}
I^{\rm edm}_{\alpha\mu\beta} & = & (-1)\int^{}_{}\frac{d^4k}{(2\pi)^4}tr_{CFD}[
     \Gamma^\rho_\beta (k;p-q)S(k_{-+})\imath\Gamma_\mu (k_{++},k_{-+}) 
     \nonumber \\
                   &   &\times S(k_{++})\Gamma^\rho_\alpha (k+q/2;p)
     S(k_{+-})]. 
\end{eqnarray}
The expanded form of the integral $I^{\rm edm}_{\alpha\mu\beta}$ 
is given in appendix B.
The EDM can now be calculated. Once again the Dirac traces are performed
and the charge and electric dipole terms projected out using the projection
operators given above. The integrals are performed and the dipole 
calculated using
\begin{eqnarray}
d & = & \,\lim_{q^2\rightarrow 0}{\cal D}(q^2)\nonumber \\
  & = & \,\lim_{q^2\rightarrow 0}P_3^{\alpha\mu\beta}
        I^{{\rm edm}}_{\alpha\mu\beta}.
\end{eqnarray}
The integral is again evaluated in the Breit frame. 

The dipole interaction between the quark and the photon is governed by the 
term
\begin{equation}
D_u\bar{u}\gamma_5\sigma^{\mu\nu}q_\nu uA_\mu+D_d\bar{d}\gamma_5\sigma^{\mu\nu}
q_\nu dA_\mu ,
\end{equation}
where the $D_u$ and $D_d$ are the electric dipole moments for the $u$ and
$d$-quarks respectively. In many models $D\propto Q$ for quarks, so the 
following form is used,
\begin{equation}
D_{\{ u,d\} } = Q_{\{ u,d\} }\ 
d_q.
\end{equation}
The perturbative result has been calculated and it gives:
\begin{equation}
d^{\rm pert.} = 3m_qd_q \mbox{  in units of e/2m}.
\end{equation}

The quark electric dipole moment has been left as a free 
parameter and so an estimate
of the size of $d_q$ is needed.
The quark electric dipole moment can be generated in various models. One
of these is the Weinberg model of three Higgs doublets \cite{Weinberg76}.
In this model the quark EDM comes about due to neutral Higgs exchange
with the quark dipole moment given by \cite{DeshpandeMa77}
\begin{equation}
d_q = \frac{eG_F}{\sqrt{2}\pi^2}m_q\sum_{i}X^\prime_iY^\prime_i
      \frac{m_q^2}{m_{Hi}^2}{\rm ln}\frac{m_q^2}{m_{Hi}^2},
\end{equation}
where the $m_{Hi}$ are the masses of the 
neutral Higgs and $X^\prime_i$ and $Y^\prime_i$ are mixings of the 
neutral Higgs. This gives a quark dipole moment of the
order of $10^{-24}$e.cm. Thus we obtain the 
results for the rho electric dipole moment,
in units of $d_q$, listed in Table~\ref{table2}.
Again the perturbative result is included for comparison.
For completeness, it is noted that the non-relativistic result is given by
$d_\rho = d_q$ e.cm.
The results from the Chappell, and Pichowsky and Lee, amplitudes
are again seen to be very close, as was the case for the magnetic moment.

\section{The Bag Model}
There are purely phenomenological models that have been used over the
years to describe the properties of hadrons. One successful and popular
model which incorporates many of the features of QCD is the MIT bag model
\cite{ChodosJaffeJohnsonThornWeisskopf74,ChodosJaffeJohnsonThorn74}. 
The MIT bag model includes relativistic, point-like, confined quarks and
provides a phenomenological description of the non-perturbative gluon 
interactions. The simplest form of the MIT model will be used, with one-gluon
exchange and centre of mass motion ignored. While these results, at least for
the magnetic moments of baryons, are well known we discuss them here to 
provide another point of comparison for the results of section~\ref{sec2}.

The starting point for the bag model is the assumption of free quarks confined
within a spherical volume of radius $R$. Solving the Dirac equation inside
the sphere results in a wavefunction given by
\cite{ChodosJaffeJohnsonThorn74,Thomas84}
\begin{equation}
\psi (r) = \frac{N(x)}{\sqrt{4\pi}}\left( \begin{array}{c}
		  \sqrt{\frac{w+m}{w}}j_0\left(\frac{xr}{R}\right)\chi \\
                  \sqrt{\frac{w-m}{w}}\imath\mbox{\boldmath 
                  $\sigma\cdot\hat{r}$}j_1\left(
                  \frac{xr}{R}\right)\chi
		                          \end{array} \right),
\end{equation}
where $w(m,R) = \sqrt{m_q^2+x^2/R^2}$, $j_0$ and $j_1$ are spherical Bessel
functions and
$x$ is the quark momentum in units
of $1/R$. The normalisation, $N(x)$, is given by
\begin{equation}
N^{-2}(x) = R^3j_0^2(x)\frac{2w(w-1/R)+m_q/R}{w(w-m)},
\end{equation}
and the eigenfrequency $x$ by
\begin{equation}
\label{eq:eigen}
\tan x = \frac{x}{1-m_qR-\left(x^2+(m_qR)^2\right)^{1/2}},
\end{equation}
arising from the boundary condition:
\begin{equation}
-\imath\mbox{\boldmath $\gamma$}\cdot\mbox{\boldmath $\hat{r}$}\psi = \psi,
\mbox{  at $r = R$}.
\end{equation}
To confine the quarks a ``pressure''
$B$ is introduced which ensures energy-momentum conservation and confinement.
Thus the total energy of the bag becomes
\begin{equation}
\label{eq:energy}
E(R) = {\textstyle \sum_{i}^{}}N_i\left( m_{qi}^2+x^2/R^2\right)^{1/2}+
       B\frac{4}{3}\pi R^3,
\end{equation}
where $N_i$ is the number of quarks or type $i$ inside the bag. The stability
of the bag also implies that
\begin{equation}
\label{eq:stability}
\frac{\partial E(R)}{\partial R} = 0.
\end{equation}
All calculations will be carried out in the rest frame of the bag and so
$E(R)$ will be equated with $m_\rho$. Equations (\ref{eq:eigen}),
(\ref{eq:energy}) and (\ref{eq:stability}) will then be used to fix the
variables $x$, $R$ and $B$.

\subsection{Magnetic Moment in the Bag Model}
The magnetic moment is given by
\begin{eqnarray}
\mbox{\boldmath $\mu$} & = & \frac{1}{2}\int_{}^{}d^3{\bf r\ r\times j}_{em} \\
      \nonumber
                       & = & \frac{1}{2}\int^{}_{}d^3{\bf r\ r}\times
                             \textstyle{\sum_{i}^{}}\left(\bar{\psi_i}
                             \mbox{\boldmath $\gamma$}\psi_i\right)Q_ie,
\end{eqnarray}
where the summation is over the quarks. This gives \cite{Thomas84,Close79}
\begin{equation}
\mbox{\boldmath $\mu_\rho$} = \mu_q\textstyle{\sum_i^{}}\mbox{\boldmath
                              $\sigma$}_iQ_i,
\end{equation}
where $\mu_q$, the magnetic dipole moment of the quark, is given by
\begin{equation}
\mu_q = \frac{e}{2m_q}\frac{1}{3}m_qR\frac{4wR+2m_qR-3}{2(wR)^2-2wR+m_qR}.
\end{equation}
Using the standard spin-flavour wavefunction of the rho yields the results,
listed in Table~\ref{table3}, and discussed in the final section of this paper.

\subsection{Electric Dipole Moment in the Bag Model}
Using the single particle wavefunction from above, and the P and CP violating
electromagnetic current
\begin{equation}
J_{\mu\nu} = -\imath D_q\bar{\psi}\gamma_5\sigma_{\mu\nu}\psi,
\end{equation}
where $\sigma_{\mu\nu} = \frac{\imath}{2}\left[\gamma_\mu,\gamma_\nu\right]$
it can be shown that 
\begin{equation}
J_{\mu\nu}F^{\mu\nu} = -D_q\bar{\psi}\mbox{\boldmath $\sigma$}\psi
                        \cdot \mbox{\boldmath $E$}+\mbox{ terms
                        involving \boldmath $B$},
\end{equation}
where $F^{\mu\nu}$ is the standard electromagnetic tensor.
Therefore \cite{CostellaMcKellar92}
\begin{eqnarray}
\mbox{\boldmath $d$}_\rho & = & 
      \int_{{\rm bag}}^{}\langle\rho\uparrow |\sum_{{\rm quarks}}
      D_q\bar{\psi}\mbox{\boldmath $\sigma$}\psi\ |\rho\uparrow\rangle 
      \nonumber \\
             & = & N^2\int_0^Rdr\ r^2
                   \left(j_0^2\left(\frac{xr}{R}\right)+\frac{1}{3}\right.
                   \left.j_1^2\left(\frac{xr}{R}\right)\right)
                   \langle\rho\uparrow |\sum_{{\rm quarks}}D_q
                   \mbox{\boldmath $\sigma$}\ |\rho\uparrow\rangle .
\end{eqnarray}
Thus, again using the rho spin-flavour wavefunction, the following 
formula for the EDM of the rho is obtained,
\begin{equation}
d_\rho = N^2\int_0^Rdr\ r^2\left(j_0^2\left(\frac{xr}{R}\right)+
        \frac{1}{3}j_1^2\left(\frac{xr}{R}
	\right)\right),\mbox{ in units of $d_q$ e cm.}
\end{equation}
This leads to the results given in Table~\ref{table4} and discussed below.

\section{Discussion}
Using a model form for the quark propagator obtained from the Dyson-
Schwinger equations, two different phenomenological rho-meson
amplitudes fitted to $f_\rho$
and $g_{\rho\pi\pi}$ and one used to describe $\rho$-$\omega$ mixing, and
a dressed quark-photon vertex, the rho dipole
moments have been calculated. 

All three non-perturbative models used, agree qualitatively,
with the amplitude and propagator defined in 
\cite{Chappell96} yielding the largest value for $\mu_\rho$
and that of \cite{MitchellTandy97} the smallest.
The Bethe-Salpeter amplitude of
\cite{PichowskyLee96} resulted in
the largest value for $d_\rho$, with that of \cite{MitchellTandy97} yielding
the smallest.
Thus
the values calculated for the rho magnetic and electric dipole moments
are fairly robust when it comes to slight changes in the forms for
both the quark propagator and the rho Bethe-Salpeter amplitude.

As seen above the inclusion of quark confinement gives greatly enhanced
values over those obtained using perturbation theory. The non-perturbative
results for the magnetic dipole moment are $\sim\!\! 29$\%--$51$\% 
bigger than the
perturbative result. While the electric dipole moment showed an even
greater increase, being $\sim\!\! 63$--$78$
times larger than the value obtained with 
perturbation theory. $d_\rho$ is linearly dependent upon the quark dipole
moment and so a different model for the quark moment could change 
the value for $d_\rho$ substantially. Conversely, an experimental
measure for $d_\rho$ would put a limit on $d_q$ and possibly rule out
some models.

The increase in the dipole moments is due to the inclusion of quark 
propagators that ensure confinement and a more accurate modelling of the
momentum distribution and finite size of the rho meson. The kernel of the 
Bethe-Salpeter equation involves the dressed quark and gluon propagators
and so includes the effects of the quark-gluon sea. The amplitudes that have
been used here can thus be said to model these effects phenomenologically.

These calculations extend the process of using the 
Dyson-Schwinger/Bethe-Salpeter framework to obtain 
hadron properties, to the rho meson.

The rho dipole moments have also been calculated in the MIT bag model. This
phenomenological model also includes confinement via the introduction of
the ``pressure'' term $B$. The bag is also assumed to incorporate the effects
of the non-perturbative gluon exchange and interaction. This relativistic,
confining model also yields values for the dipole moments greater than that of
perturbation theory, as was the case for the results using the DSE/BSE 
framework, {\it i.e.,} the results from both approaches agree qualitatively.
The magnetic dipole moment is about $\sim\!\! 10$\% bigger than the 
perturbative result, smaller than any of the results obtained in the 
impulse approximation. The electric dipole 
calculation was $\sim\!\! 6$\% bigger
than the largest result using the Dyson-Schwinger/Bethe-Salpeter equations and
$\sim\!\! 83$ times bigger than the perturbative result. 
Again the inclusion
of confinement has led to an enhancement in the dipole moments of the rho.
Indeed, one could suggest that for the calculation of the magnetic and 
electric dipole moments, the bag model gives a reasonable estimate of the
effects of confinement.

The rho meson has been used in this study, not because there is any
 realistic possibility of measuring the electromagnetic moments we have
 calculated, but to provide some insight into the effects of QCD and
 confinement on calculations of these moments in the hope of obtaining
 insight into the effects these aspects of quark dynamics may have on 
 estimates of the electromagnetic moments of hadrons 
 \cite{TupperMcKellarWarner89,HeMcKellarPakvasa89}. 

\acknowledgements

M.B.H. wishes to thank C.D. Roberts for helpful discussions
and for pointing out the work of I. Chappell. M.B.H. also wishes to
acknowledge the financial support given by the Australian Postgraduate Award.
This work was supported in part by the Australian Research Grants Committee,
and the Australian Department of Industry, Science and Tourism.

\appendix

\section{Integral for the Magnetic Dipole Moment}
With the definitions given above the integral $I_{\alpha\mu\beta}$ becomes
(after colour and flavour traces)
\begin{eqnarray}
I_{\alpha\mu\beta} & = & \frac{2\imath N_c}{N_\rho^2}\int^{}_{}
     \frac{d^4k}{(2\pi)^4}\Gamma_\rho (k)\Gamma_\rho (k+q/2)\; tr_D\left[
     \frac{}{}\right. \nonumber \\
                   &   & \gamma_\beta\skmp\gamma_\mu\skpp\gamma_\alpha
     \skpm T_1 +\gamma_\beta\skmp\gamma_\mu\gamma_\alpha T_2 \nonumber \\
                   &   & +\gb\skmp\ga\skpm\kopm T_3+\gb\skmp\skpp\ga\kopm T_4
     \nonumber \\
                   &   & +\gb\skpp\ga\skpm\kopm T_5+\gb\ga\kopm T_6 \nonumber 
     \\
                   &   & +\gb\skmp\ga\skpm\kopm T_7+\gb\skmp\skpp\ga\kopm T_8
     \nonumber \\
                   &   & +\gb\gm\ga\skpm T_9+\gb\gm\skpp\ga T_{10} \nonumber \\
                   &   & +\gb\skpp\ga\skpm\kopm T_{11}+\gb\ga\kopm T_{12}
     \nonumber \\
                   &   & +\gb\skmp\ga\skpm\kopm T_{13}+\gb\skmp\skpp\ga\kopm
     T_{14} \nonumber \\
                   &   & +\left.\gb\skpp\ga\skpm\kopm T_{15}+\gb\ga\kopm T_{16}
                         \frac{}{}\right],
\end{eqnarray}
where the terms with an odd number of $\gamma$ matrices have been dropped.
The $T_{1-16}$ are defined as follows:
\begin{eqnarray}
T_1  & = & \imath (V_1+V_4)\svkmp\svkpp\svkpm, \nonumber \\
T_2  & = & -\imath (V_1+V_4)\svkmp\sskpp\sskpm, \nonumber \\
T_3  & = & 2\imath k_{++}^2(V_2+V_5)\svkmp\svkpp\svkpm, \nonumber \\
T_4  & = & -2\imath (V_2+V_5)\svkmp\sskpp\sskpm, \nonumber \\
T_5  & = & 2\imath k_{-+}^2(V_2-V_5)\svkmp\svkpp\svkpm, \nonumber \\
T_6  & = & -2\imath k_{-+}^2(V_2-V_5)\svkmp\sskpp\sskpm, \nonumber \\
T_7  & = & -2 V_3\svkmp\sskpp\svkpm \nonumber, \\
T_8  & = & -2 V_3\svkmp\svkpp\sskpm \nonumber, \\
T_9  & = & -\imath (V_1+V_4)\sskmp\sskpp\svkpm, \nonumber \\
T_{10} & = & -\imath (V_1+V_4)\sskmp\svkpp\sskpm, \nonumber \\
T_{11} & = & -2\imath (V_2+V_5)\sskmp\sskpp\svkpm, \nonumber \\
T_{12} & = & -2\imath k_{++}^2(V_2+V_5)\sskmp\svkpp\sskpm, \nonumber \\
T_{13} & = & -2\imath (V_2-V_5)\sskmp\sskpp\svkpm, \nonumber \\
T_{14} & = & -2\imath (V_2-V_5)\sskmp\svkpp\sskpm, \nonumber \\
T_{15} & = & -2V_3\sskmp\svkpp\svkpm, \nonumber \\
T_{16} & = & 2V_3\sskmp\sskpp\sskpm,
\end{eqnarray}
with,
\begin{eqnarray}
V_1  & = & \frac{A(k_{++}^2)+A(k_{-+}^2)}{2}, \nonumber \\
V_2  & = & \frac{A(k_{++}^2)-A(k_{-+}^2)}{2(k_{++}^2-k_{-+}^2)}, \nonumber \\
V_3  & = & \frac{-\imath (B(k_{++}^2)-B(k_{-+}^2))}{k_{++}^2-k_{-+}^2},
           \nonumber \\
V_4  & = & \frac{-\imath (k_{++}^2-k_{-+}^2)(A(k_{++}^2)-A(k_{-+}^2))}{
           2d(k_{++},k_{-+})}, \nonumber \\
V_5  & = & \frac{-(A(k_{++}^2)-A(k_{-+}^2))}{2d(k_{++},k_{-+})}.
\end{eqnarray}

\section{Integral for the Electric Dipole Moment}
The integral $I^{\rm edm}_{\alpha\mu\beta}$ becomes (after colour and flavour
traces)
\begin{eqnarray}
I^{\rm edm}_{\alpha\mu\beta} & = & \frac{2\imath N_c}{N_\rho^2}\int^{}_{}
    \frac{d^4k}{(2\pi )^4}\Gamma_\rho (k)\Gamma_\rho (k+q/2)tr_D\left[
    \frac{}{}\right. \nonumber \\
                             &   & \gb\skmp\gm\skpp\ga\skpm T_1+
    \gb\skmp\gm\ga T_2 \nonumber \\
                             &   & +\gb\skmp\ga\skpm\kopm T_3+
    \gb\skmp\skpp\ga\kopm T_4 \nonumber \\
                             &   & +\gb\skmp\ga\skpm\kopm T_5+
    \gb\ga\kopm T_6 \nonumber \\
                             &   & +\gb\skmp\skpp\ga\kopm T_7+
    \gb\skmp\ga\skpm\kopm T_8 \nonumber \\
                             &   & +\gb\skmp\gamma_5\sigma_{\mu\rho}q^\rho
    \skpp\ga T_9+
    \gb\skmp\gamma_5\sigma_{\mu\rho}q^\rho\ga\skpm T_{10} \nonumber \\
  			     &   & +\gb\gm\skpp\ga T_{11}+
    \gb\gm\ga\skpm T_{12} \nonumber \\
			     &   & +\gb\ga\kopm T_{13}+
    \gb\skpp\ga\skpm\kopm T_{14} \nonumber \\
			     &   & +\gb\skmp\skpp\ga\kopm T_{15}+
    \gb\skmp\ga\skpm\kopm T_{16} \nonumber \\
			     &   & +\gb\skpp\ga\skpm\kopm T_{17}+
    \gb\ga\kopm T_{18} \nonumber \\
			     &   & +\gb\gamma_5\sigma{\mu\rho}q^\rho\skpp
    \ga\skpm T_{19}+
    \gb\gamma_5\sigma_{\mu\rho}q^\rho\ga T_{20}
    \left.\frac{}{}\right].
\end{eqnarray}
Again dropping terms with an odd number of $\gamma$ matrices.
The $T_{1-20}$ are defined as follows:
\begin{eqnarray}
T_1 & = & \imath (V_1+V_4)\svkmp\svkpp\svkpm \nonumber \\
T_2 & = & -\imath (V_1+V_4)\svkmp\sskpp\sskpm \nonumber \\
T_3 & = & 2\imath\kpp^2(V_2+V_5)\svkmp\svkpp\svkpm \nonumber \\
T_4 & = & -2\imath (V_2+V_5)\svkmp\sskpp\sskpm \nonumber \\
T_5 & = & 2\imath\kmp^2(V_2-V_5)\svkmp\svkpp\svkpm \nonumber \\
T_6 & = & -2\imath\kmp^2(V_2-V_5)\svkmp\sskpp\sskpm \nonumber \\
T_7 & = & -2V_3\svkmp\svkpp\sskpm \nonumber \\
T_8 & = & -2V_3\svkmp\sskpp\svkpm \nonumber \\
T_9 & = & -V_6\svkmp\svkpp\sskpm \nonumber \\
T_{10} & = & -V_6\svkmp\sskpp\svkpm \nonumber \\
T_{11} & = & -\imath (V_1+V_4)\sskmp\svkpp\sskpm \nonumber \\
T_{12} & = & -\imath (V_1+V_4)\sskmp\sskpp\svkpm \nonumber \\
T_{13} & = & -2\imath\kpp^2(V_2+V_5)\sskmp\svkpp\sskpm \nonumber \\
T_{14} & = & -2\imath (V_2+V_5)\sskmp\sskpp\svkpm \nonumber \\
T_{15} & = & -2\imath (V_2-V_5)\sskmp\svkpp\sskpm \nonumber \\
T_{16} & = & -2\imath (V_2-V_5)\sskmp\sskpp\svkpm \nonumber \\
T_{17} & = & -2V_3\sskmp\svkpp\svkpm \nonumber \\
T_{18} & = & 2V_3\sskmp\sskpp\sskpm \nonumber \\
T_{19} & = & -V_6\sskmp\svkpp\svkpm \nonumber \\
T_{20} & = & V_6\sskmp\sskpp\sskpm,
\end{eqnarray}
where $V_1$\dots$V_5$ are defined as before and
\begin{equation}
V_6 = -\imath d_q, \mbox{ }d_q\mbox{ the quark dipole moment.}
\end{equation}

\newpage

\begin{table}[h]
\caption{Magnetic Dipole Moment $\mu$ in units of e/2m}
\label{table1}
\begin{center}
\begin{tabular}{|c|c|} \hline
Rho Amplitude/Quark Propagator & $\mu$ \\ \hline
Chappell \cite{Chappell96} & 3.01 \\
Pichowsky and Lee \cite{PichowskyLee96} & 2.72 \\
Mitchell and Tandy \cite{MitchellTandy97} & 2.57 \\
Perturbative Result \cite{Pallaghy96} & 2.00  \\ \hline
Non-Relativistic SU(6) \cite{Hecht94}, constituent quark mass 
from \cite{Close79} & 2.26 \\ \hline
\end{tabular}
\end{center}
\end{table}
\newpage

\begin{table}[h]
\caption{Electric Dipole Moment $d_\rho$ in units of $d_q$ e.cm.}
\label{table2}
\begin{center}
\begin{tabular}{|c|c|} \hline
Rho Amplitude/Quark Propagator & $d_\rho$ \\ \hline
Chappell \cite{Chappell96} & $0.743$ \\
Pichowsky and Lee \cite{PichowskyLee96} & $0.779$ \\
Mitchell and Tandy \cite{MitchellTandy97} & $0.627$ \\
Perturbative Result \cite{Pallaghy96}, $m_q$ from 
\cite{Chappell96,PichowskyLee96}
& $0.010$ \\ \hline
\end{tabular}
\end{center}
\end{table}
\newpage

\begin{table}[h]
\caption{Magnetic Dipole Moment $\mu$ in the Bag Model in units of e/2m}
\label{table3}
\begin{center}
\begin{tabular}{|c|c|} \hline
Mass of Quark & $\mu$ \\ \hline
$m_q$ from \cite{Chappell96,PichowskyLee96} & 2.21 \\ \hline
Chappell \cite{Chappell96} & 3.01 \\
Pichowsky and Lee \cite{PichowskyLee96} & 2.72 \\
Mitchell and Tandy \cite{MitchellTandy97} & 2.57 \\ \hline
\end{tabular}
\end{center}
\end{table}
\newpage

\begin{table}[h]
\caption{Electric Dipole Moment in the Bag Model in units of $d_q$ e cm.}
\label{table4}
\begin{center}
\begin{tabular}{|c|c|} \hline
Mass of Quark & $d_\rho$ \\ \hline
$m_q$ from \cite{Chappell96,PichowskyLee96} & $0.828$ \\ \hline
Chappell \cite{Chappell96} & $0.743$ \\
Pichowsky and Lee \cite{PichowskyLee96} & $0.779$ \\
Mitchell and Tandy \cite{MitchellTandy97} & $0.627$ \\ \hline
\end{tabular}
\end{center}
\end{table}
\newpage

\begin{center}
FIG. 1. The dressed $\rho\rho\gamma$ Vertex
\end{center}
\newpage

\begin{figure}[h]
\hspace{5.5cm}
\psfig{file=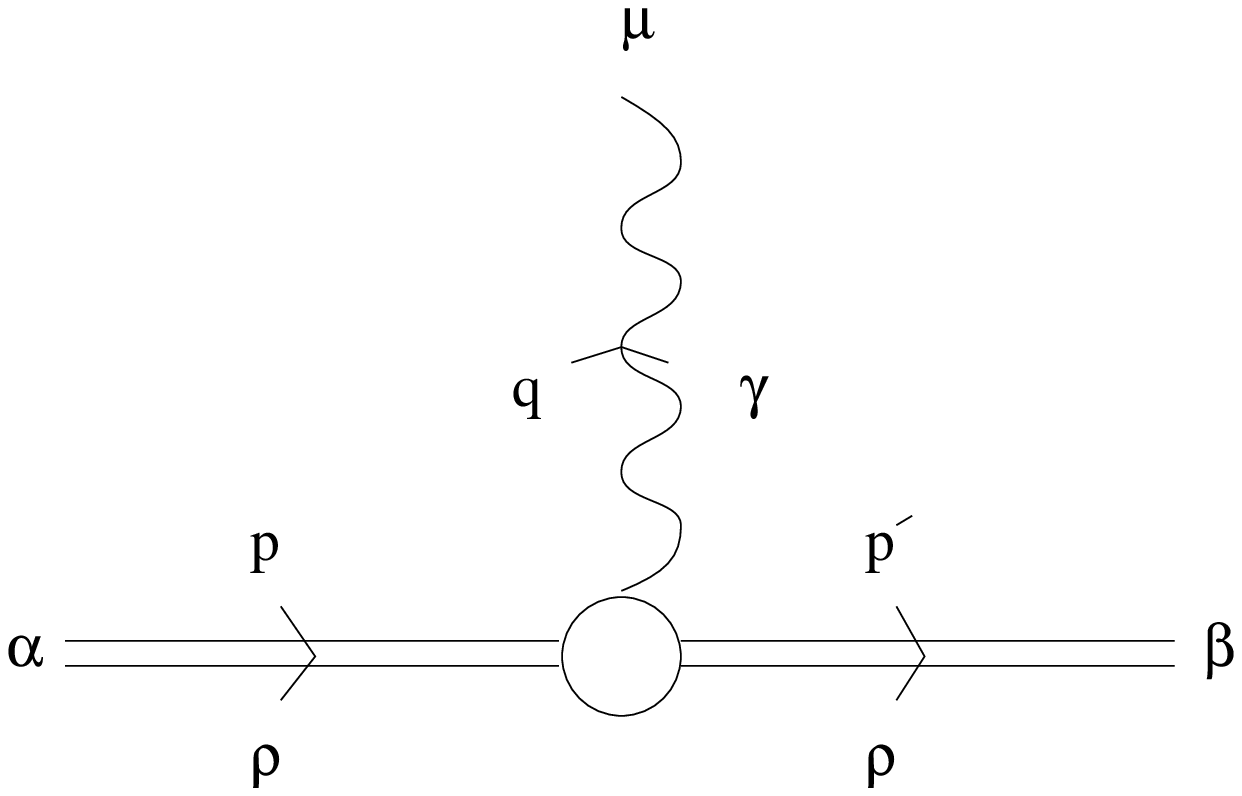,width=6cm,height=3cm}
\clearpage
\label{figure1}
\end{figure}
\newpage

\begin{center}
FIG. 2. The quark contribution to the dipole moments of the rho,
	with $S$ the dressed quark propagators, $\Gamma_\mu$ the dressed
	quark-photon vertex and $\Gamma^\rho$ the rho-quark vertex.
\end{center}
\newpage

\begin{figure}[h]
\hspace{2.5cm}
\psfig{file=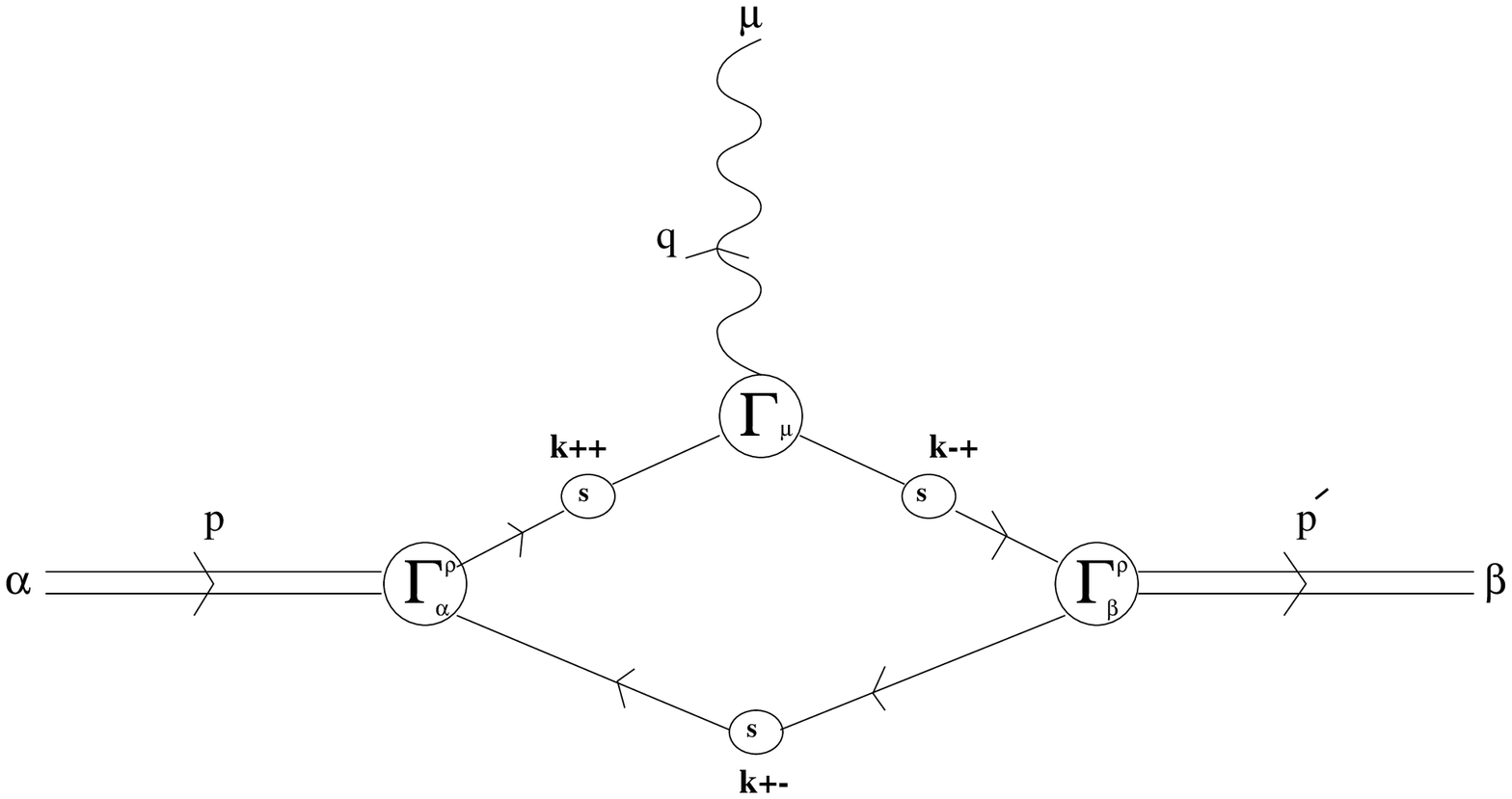,width=12cm,height=6cm}
\label{figure2}
\end{figure}
\end{document}